\documentstyle[aps,twocolumn,epsfig,amssymb]{revtex}
\def\ave#1{\langle #1\rangle}

\newcommand{\op}[1]{\hat{#1}}

\newcommand{\bra}[1]{\langle #1|}
\newcommand{\ket}[1]{|#1\rangle}
\newcommand{\braket}[2]{\langle #1|#2\rangle}

\newcommand{\tr}{{\,\rm tr\,}}
\newcommand{\ul}[1]{{\underline #1}}
\begin{document}
\title{On general relation between quantum ergodicity and fidelity of quantum dynamics
}
\author{Toma\v z Prosen}
\address{Physics Department, Faculty of Mathematics and Physics, 
University of Ljubljana, Slovenia}
\date{\today}
\draft
\maketitle
\begin{abstract}
General relation is derived which expresses the fidelity of quantum dynamics, 
measuring the stability of time evolution to small static variation in the 
hamiltonian, in terms of ergodicity of an observable
generating the perturbation as defined by its time correlation function.
Fidelity for {\em ergodic} dynamics is predicted to decay 
{\em exponentially} on time-scale $\propto\delta^{-2}$, 
$\delta\sim$ strength of perturbation, whereas faster, typically 
{\em gaussian} decay on shorter time scale $\propto \delta^{-1}$ 
is predicted for {\em integrable}, or generally 
{\em non-ergodic} dynamics. This surprising result is demonstrated in 
quantum Ising spin-$1/2$ chain periodically {\em kicked} 
with a tilted magnetic field where we find finite parameter-space 
regions of non-ergodic and non-integrable 
motion in thermodynamic limit. 
\end{abstract}

\pacs{PACS number: 05.45.-a, 03.65.Yz, 75.10.Jm}

The quantum signatures of various types of classical motion, ranging from integrable to
ergodic, mixing and chaotic, are still lively debated issues (see e.g.
\cite{Qchaos}).
Most controversial is the absence of exponential sensitivity to variation
of initial condition in quantum mechanics which prevents direct definition of quantum chaos
\cite{casati}. However, there is an alternative concept which can be used in
classical as well as in quantum mechanics \cite{peres}: One can study 
stability of motion with respect to small variation in the Hamiltonian. 
Clearly, in classical mechanics this concept, when
applied to individual trajectories, is equivalent to sensitivity to initial conditions. 
Integrable systems with regular orbits are stable against small variation in the hamiltonian 
(the statement of KAM theorem), wheres for chaotic orbits varying the
hamiltonian has similar effect as varying the initial condition: exponential divergence of two
orbits for two nearby chaotic hamiltonians. 

The quantity of the central interest here is the {\em fidelity} 
of quantum motion.
Consider a unitary operator $U$ being either (i) a short-time propagator, 
or (ii) a Floquet map 
$U=\op{\cal T}\exp(-i\int_0^p d\tau H(\tau)/\hbar)$ of 
(periodically time-dependent) Hamiltonian 
$H$ ($H(\tau+p)=H(\tau)$), or (iii) a quantum Poincar\' e map. 
The influence of a small perturbation to the unitary evolution, 
which is generated by a hermitean operator $A$, 
$U_\delta = U\exp(-i A \delta)$, $\delta$ being a small parameter, 
is described by the overlap $\braket{\psi_\delta(t)}{\psi(t)}$ measuring 
the Hilbert space distance between exact and perturbed time evolution 
from the same initial pure state 
$\ket{\psi(t)} = U^t\ket{\psi}$, 
$\ket{\psi_\delta(t)} = U^t_\delta\ket{\psi}$, where {\em integer} $t$ is a
discrete time (in units of period $p$)\cite{foot0}.
This defines the {\em fidelity} 
\begin{equation}
F(t) = \ave{U_\delta^{-t} U^t},
\label{eq:deffid}
\end{equation}
where the average is performed either over a 
fixed pure state $\ave{.} = \bra{\psi}.\ket{\psi}$,
or, if convenient, as a uniform average over all possible initial 
states 
$\ave{.} = (1/{\cal N})\tr (.)$, ${\cal N}$ being the Hilbert space dimension.
The quantity $F(t)$ has already raised considerable interest, though under
different names and interpretations: First, it has been proposed by Peres\cite{peres} 
as a measure of stability of quantum motion. Second, it is the {\em Loschmidt echo} 
measuring the {\em dynamical irreversibility of quantum phases}, used e.g. in spin-echo 
experiments\cite{Usaj} where one is interested in the overlap between the initial state 
$\ket{\psi}$ and a state $U_\delta^{-t} U^t\ket{\psi}$ obtained 
by composing forward time evolution, imperfect time inversion with a 
residual interaction described by the operator $A\delta$, and backward time evolution.
Third, the fidelity has become a standard measure characterizing the loss of phase coherence in 
quantum computation\cite{qcomp}. Fourth, it was used to characterize 
``hypersensitivity to perturbation'' in related studies\cite{SC96}, 
though in different contexts of stochastically time-dependent perturbation.

The main result of this paper is a relation of the fidelity to 
ergodic properties of quantum dynamics, more precisely to the time 
autocorrelation function of the generator of the perturbation $A$. 
Quantum dynamics of finite and bound systems has always a {\em discrete 
spectrum} since the effective Hilbert space dimension ${\cal N}$ is finite, 
hence it is {\em non-ergodic} and {\em non-mixing}\cite{qmix,ktv}: 
time correlation functions have fluctuating tails of order $\sim 1/{\cal N}$. 
In order to reach genuine complexity of quantum motion with possibly 
continuous spectrum one has to enforce ${\cal N}\to\infty$ by considering 
one of the following two limits: quasi-classical limit of effective 
Planck's constant $\hbar\to 0$, or thermodynamic limit (TL) of number of 
particles, or size $L\to\infty$. 
Our result is surprising in the sense that it predicts the {\em average} 
fidelity to exhibit exponential decay on a time scale $\propto \delta^{-2}$ 
for {\em ergodic systems} (i.e. such that the integrated time auto-correlation of
$A$ is finite), but much faster, typically gaussian decay on a shorter time 
scale $\propto \delta^{-1}$ for integrable and  
general non-ergodic systems (i.e. such that time averaged auto-correlation of
$A$ is non-vanishing). 
Our theory on fidelity is very general and can be extended to any perturbed unitary 
evolution, either in quantum, quasi-classical, or even classical (Liouvillian) context.
In this paper we apply it to the {\em quantum many-body} problem in TL, in particular in
the {\em Kicked Ising model} (KI), namely the Ising spin $1/2$ chain periodically kicked with 
a tilted homogeneous magnetic field. KI is particularly interesting since it possesses 
parameter-space regions with positive measure of {\em non-ergodic} behavior in TL 
surrounding the integrable cases\cite{ki} of vanishing measure, which is an additional 
evidence for a conjecture\cite{ktv} on existence of intermediate, non-integrable and 
non-ergodic quantum motion of disorderless interacting many-body systems in TL. 

We start by rewriting the fidelity (\ref{eq:deffid}) in terms of Heisenberg
evolution of the perturbation $A_t:=U^{-t}AU^t$
\begin{equation}
F(t) = \ave{e^{iA_{0}\delta}e^{iA_{1}\delta}\cdots e^{iA_{t-1}\delta}} = 
\op{\cal T}\ave{\prod_{t'=0}^{t-1}\exp(i A_{t'}\delta)}
\label{eq:prodfid}
\end{equation}
which is achieved by $t$ insertions of the unity $U^{-t'} U^{t'}$ and recognizing
$U^{-(t'-1)}U^\dagger_\delta U^{t'} = \exp(i\delta A_{t'-1})$. 
$\op{\cal T}$ is a left-to-right time ordering.
Next we make an expansion in $\delta$ expressing the fidelity in terms of correlation functions
\begin{equation}
F(t) = 1 + \sum_{m=1}^\infty \frac{i^m \delta^m}{m!}
\op{\cal T}\!\!\!\!\!\sum_{t_1,t_2\ldots t_m=0}^{t-1} \ave{A_{t_1}A_{t_2}\cdots A_{t_m}}.
\label{eq:serfid}
\end{equation}
Being interested mainly in the absolute value $|F(t)|$ we will in the following choose 
perturbations with vanishing first moment 
$a:=(1/t)\sum_{t'=0}^{t-1}\ave{A_{t'}}=0$ so that the series
(\ref{eq:serfid}) starts at $m=2$, since a shift by a multiple 
of unity $A\to A-a 1$ simply rotates the fidelity $F(t) \to \exp(-ia\delta)F(t)$. 
On the other hand, we can eliminate not only the first, $m=1$, 
but all {\em odd} orders in the expansion
(\ref{eq:serfid}) by considering the {\em symmetrized} fidelity $F(t) = 
\ave{U^{-t}_{\delta/2}U^t_{-\delta/2}}$.
To second order in $\delta$ we have
\begin{equation}
F(t) = 1 - \frac{\delta^2}{2} \sum_{t'=-t}^{t}(t-|t'|)C_A(t') + {\cal O}(\delta^3),
\label{eq:quadfid}
\end{equation}
where it is assumed that 2-point time correlation function is homogeneous 
$C_A(t'-t):=\ave{A_t A_{t'}}$, 
as is the case for uniform average over initial states $\ave{.}=\tr(.)/{\cal N}$. 
Eq.~(\ref{eq:quadfid}) reveals a simple general rule: the stronger correlation decay, 
the slower is decay in fidelity, and vice versa. 
Below we discuss two different cases in the limit ${\cal N}\to\infty$:

{\em I. Ergodicity and fast mixing}. Here we 
assume that $C_A(t)\to 0$ sufficiently fast that the total sum converges, 
$S_A:=(1/2)\sum_{t=-\infty}^\infty C_A(t)$, $|S_A| < \infty$. 
For times $t$ much larger than the so-called {\em mixing time scale} 
$t\gg t_{\rm mix}$ 
which effectively characterizes the correlation decay, 
e.g. $t_{\rm mix} = \sum_t |t C_A(t)|/\sum_t |C_A(t)|$, it follows that the fidelity
drops linearly in time $F_{\rm e}(t) = 1 - t/\tau_{\rm e} + {\cal O}(\delta^3)$ on a scale
\begin{equation}
\tau_{\rm e} = S_A^{-1} \delta^{-2}.
\label{eq:taue}
\end{equation}
In order to show even stronger result we further assume fast mixing with respect to product observables 
$B_{t t'} = A_t A_{t'}$ with $\ave{B_{t t'}} = C_A(t'-t)$, of order $k\ge 2$, namely 
$\ave{B_{t_1 t_2} B_{t_3 t_4}\cdots B_{t_{2k-1} t_{2k}}} \to 
\prod_{j=1}^k \ave{B_{t_{2j-1} t_{2j}}}$ as $t_1,t_2,\ldots$ are ordered and 
$t_{2j+1}-t_{2j}\to\infty$. Therefore, the leading contribution for large $t$ to each $m$-term of
(\ref{eq:serfid}) comes from sequences $(t_1,t_2,\ldots t_m)$ where consecutive pairs 
$(t_{2j-1},t_{2j})$ are close to each other, 
$t_{2j}-t_{2j-1} \stackrel{<}{\sim} t_{\rm mix}$. 
Since for odd $m$ time indices cannot be paired these terms should vanish asymptotically (as $t\to\infty$) 
relatively to even $m$ terms. Thus we can evaluate $(2k-1)!!$ equivalent even $m=2k$ terms in Eq.~(\ref{eq:serfid}) as $k$-tuple of independent sums over $t'_j=t_{2j}-t_{2j-1}$ giving, for 
$t\gg t_{\rm mix}$
\begin{equation}
F_{\rm e}(t) = \sum_{k=0}^\infty \frac{(-1)^k(2k-1)!!2^k\delta^{2k}S^k_A}{(2k)!} = 
\exp(-t/\tau_{\rm e}).
\label{eq:fide}
\end{equation}
Note that formulae (\ref{eq:taue},\ref{eq:fide}) remain valid in a more general case of inhomogeneous
time correlations where one should take $S_A:=\lim_{t\to\infty}(1/t)\sum_{t,t'=0}^\infty
\ave{B_{tt'}}$. 

{\em II. Non-ergodicity}. Here we assume that auto-correlation function of the perturbation does 
not decay asymptotically but has a non-vanishing time-average,
$D_A := \lim_{t\to\infty}(1/t)\sum_{t'=0}^{t-1}C_A(t')$,
though the first moment is vanishing $\ave{A}=0$. For times $t$ larger than the {\em averaging time}
$t_{\rm ave}$ in which a finite time average effectively relaxes into the stationary value $D_A$, we can
write fidelity to second order which decays quadratically in time,
$F_{\rm ne}(t) = 1 - (1/2)(t/\tau_{\rm ne})^2 + {\cal O}(\delta^2)$, on a scale
\begin{equation}
\tau_{\rm ne} = D_A^{-1/2} \delta^{-1}.
\label{eq:taune}
\end{equation}
More general result can be formulated in terms of a time averaged operator
$\bar{A}:=\lim_{t\to\infty}(1/t)\sum_{t'=0}^{t-1}A_{t'}$, 
namely for $t\gg t_{\rm ave}$ Eq.~(\ref{eq:serfid}) can be rewritten as
\begin{equation}
F_{\rm ne}(t) = 1 + \sum_{m=2}^\infty \frac{i^m \delta^m t^m}{m!}\ave{\bar{A}^m} = 
\ave{\exp(i\bar{A} \delta t)}.
\label{eq:fidne}
\end{equation}
Global behavior of $F_{\rm ne}(t)$ for non-ergodic systems, where higher 
$m$-terms of (\ref{eq:serfid}) become important, depends generally on the full 
sequence of moments $\ave{\bar{A}^m}$. We argue below,
by giving an example of spin $1/2$ chains, that there are large classes of 
perturbing operators where these moments can be shown to possess normal gaussian 
behavior, yielding Eq.~(\ref{eq:fidne2}). Non-ergodic behavior is certainly 
present for generic observables in {\em completely integrable
systems} where a sequence of conservation laws can be used to estimate the 
time-averaged correlator
$D_A$ \cite{zotos}, but we wish to make a stronger statement, namely that there is
a generic regime of intermediate dynamics in non-integrable systems displaying 
non-ergodic behavior\cite{ktv}. 

Let us now apply our theory to quantum spin-$1/2$ chains described by Pauli 
operators $\sigma^{xyz}_j$ on a periodic lattice of size $L$, $j+L\equiv j$, 
acting on a Hilbert space of dimension ${\cal N}=2^L$, fix the
average $\ave{.}=\tr(.)/{\cal N}$, and assume that our Floquet-operator $U$ is 
{\em translationally invariant} (TI) on a lattice. It is useful to introduce a set of 
{\em local} TI observables
$Z_{\ul{s}}=L^{-1/2}\sum_j \sigma^{s_0}_j\sigma^{s_1}_{j+1}
\cdots\sigma^{s_n}_{j+n}$, of {\em order} $n \ll L$, where $\ul{s} = [s_0,s_1\ldots s_n]$, 
$s_0,s_n\in\{x,y,z\}$, $s_j\in\{0,x,y,z\},1\le j\le n-1$, and $\sigma^0_j:=1$.  
Using $\ave{\sigma^s_j \sigma^{r}_{k}} = \delta_{j,k}\delta_{s,r}$ 
one may derive a contraction formula 
$$
\ave{Z_{\ul{s}_1}Z_{\ul{s}_2}\cdots Z_{\ul{s}_{2k}}} = 
\!\!\!\sum_{{\rm all\;pairings}}^{\cup\{\alpha,\beta\}=\{1\ldots 2k\}}\!\!\prod_{\alpha,\beta}
\delta_{\ul{s}_\alpha,\ul{s}_\beta} + {\cal O}(L^{-1}),
$$
while for odd number 
$\ave{Z_{\ul{s}_1}Z_{\ul{s}_2}\cdots Z_{\ul{s}_{2k+1}}} = {\cal O}(L^{-1})$, hence
$Z_{\ul{s}}$ become independent {\em gaussian} field variables in TL depending on a multi-index $\ul{s}$
of variable but finite length. Therefore, any TI {\em pseudo-local} (PL) observable $A$, 
having by definition\cite{ktv} $l^2$-expansion 
in the basis $Z_{\ul{s}}$ (when $L=\infty$), namely 
$A=\sum_{\ul{s}}a_{\ul{s}}Z_{\ul{s}}$, 
$\ave{A^2} = \sum_{\ul{s}} |a_{\ul{s}}|^2 < \infty$, possesses normal gaussian moments 
$\ave{A^{2k}} = (2k-1)!!\ave{A^2}^k (1+{\cal O}(L^{-1}))$.
Further, for a general TI PL observable $A$, its time average $\bar{A}$ is also TI PL, since it can 
be formally expanded in terms of $Z_{\ul{s}}$ due to construction of $\bar{A}$, 
and such expansion is $l^2$ since $\ave{\bar{A}^2} = \ave{\bar{A}A} = 
D_A < \ave{A^2}$ \cite{foot1}. 
However, for a more general non-TI PL observable $A$, i.e. such that its 
{\em linear projection} to the space of TI observables 
$(1/L)\sum_{n=0}^{L-1} A|_{\vec{\sigma}_j\to\vec{\sigma}_{j+n}}$ is PL,
one cannot generally show that $\bar{A}$ is TI PL although we believe that this is a 
typical situation, which we can prove in two cases: (i) If the spectrum of propagator $U$ is
non-degenerate (for any finite $L$), then the matrix of $\bar{A}$ is diagonal in the eigenbasis 
of $U$ and $\bar{A}$ is TI due to Bloch theorem. (ii) If the system is integrable
having a complete set of TI PL conservation laws $Q_n,n=1,2\ldots$ in the sense that
$\{Q_n\}$ is a complete set of eigenvectors of the 
Heisenberg map $\op{\cal U}A=U^\dagger A U$
for eigenvalue $1$ then the time average is a projection
$\bar{A} = \sum_n \ave{Q_n A}Q_n$ (assuming that $\ave{Q_n Q_m}=\delta_{nm}$)
which is TI PL. This is the case for KI model studied below.
Finally, assuming either (i), (ii), or simply TI PL perturbation $A$, we find that moments 
of time-average $\bar{A}$ are gaussian $\ave{\bar{A}^{2k}} = (2k-1)!! D_A^k (1+{\cal O}(L^{-1}))$. 
Summing up the formula (\ref{eq:fidne}) produces gaussian decay 
\begin{equation}
F_{\rm ne}(t) = \exp\left(-(t/\tau_{\rm ne})^2/2\right),
\label{eq:fidne2}
\end{equation}
for $t\gg t_{\rm ave}$, on a time scale (\ref{eq:taune}), which can be 
computed in a typical integrable situation (ii) as shown bellow.

Few remarks on the case of finite dimension ${\cal N}<\infty$ are in order:
(1) $F(t)$ will then start fluctuating around zero with 
magnitude $F_{\rm fluct} = {\cal N}^{-1/2}$ for {\em very long times} 
$t > t^*({\cal N})$ where the time scale $t^*({\cal N})$ is determined from 
the condition $F(t^*)|_{{\cal N}=\infty} = {\cal N}^{-1/2}$.
(2) $F(t)$ decays all the way down to ${\cal N}^{-1/2}$ only 
for a {\em typical} or {\em random} initial state $\ket{\psi}$ with  
$\sim{\cal N}$ non-vanishing components when expanded in the eigenbasis of $U$,
or for an average over $\ket{\psi}$. 
If on the other hand one considers the initial state which, when expanded
either in the eigenbasis of $U$ or of $U_\delta$, contains essentially only 
few, say $m$ dominating components, like the {\em regular} coherent state of 
Peres\cite{peres}, then $F(t)$ is a quasi-periodic function with $m$ 
small frequencies $\propto\delta$ and amplitudes $\sim 1/m$.
(3) Even in asymptotically ergodic situation 
the correlation $C_A(t)$ has a plateau for finite ${\cal N}$,
which can be estimated using random matrix model for the propagator $U^t$ as 
$D_A \sim D_A^*({\cal N}):=c_A/{\cal N}$ where $c_A$ is some constant with respect to
${\cal N}$.
The non-vanishing correlation plateau gives a  dominant 
contribution to Eq.~(\ref{eq:quadfid}) resulting in a quadratic (or gaussian) decay
of $F(t)$ as soon as $\tau_{\rm e} > S_A\vert_{{\cal N}=\infty}/D_A^*$,
i.e. when $\delta < \delta_{\rm p}({\cal N}) := S_A^{-1}c_A^{1/2}{\cal N}^{-1/2}$.
This {\em perturbative} regime of very small perturbation strength, existing for finite 
${\cal N}$ only, is consistent with the first order perturbation expansion of eigenstates
of $U_\delta$ in terms of the eigenbasis of $U$ \cite{cerruti}.

\begin{figure}
\vspace{-4mm}
\psfig{figure=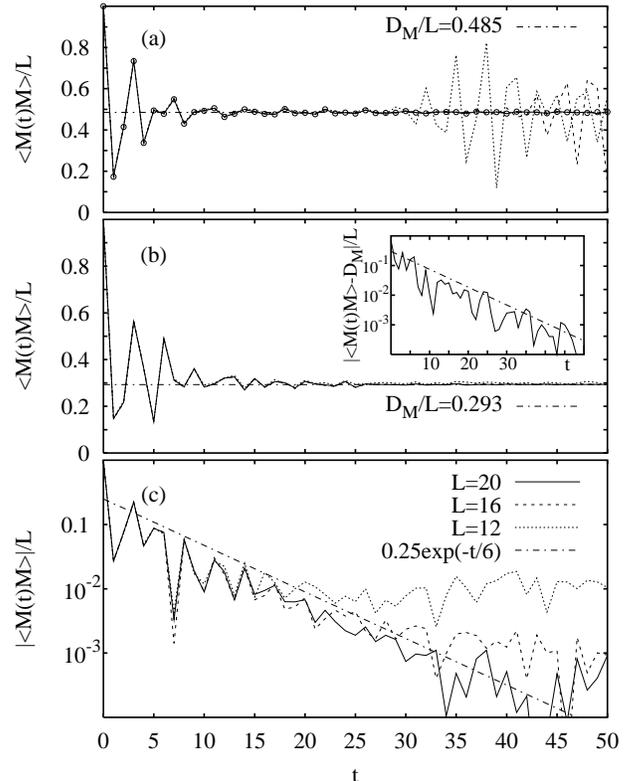,width=4.3in,angle=-90}
\caption{
Correlation decay for three cases of KI: (a) integrable $h_z=0$, (b) intermediate $h_z=0.4$, and (c)
ergodic $h_z=1.4$, for different sizes $L=20,16,12$ (solid-dotted connected curves, 
almost indistinguishable in (a,b)). Circles (a)
show exact $L=\infty$ result. Chain lines are theoretical/suggested asymptotics (see text). 
}
\label{fig:1}
\end{figure}

Consider an example of KI model with the hamiltonian
\begin{equation}
H_{\rm KI}(t) = \sum_{j=0}^{L-1} \left\{
J_z \sigma^z_j\sigma^z_{j+1} + \delta_p(t)(h_x \sigma^x_j 
+ h_z \sigma^z_j)\right\}
\label{eq:ham}
\end{equation}
where $\delta_p(t)=\sum_m\delta(t-m p)$, with a Floquet-map
$U=\exp(-iJ_z\sum_j \sigma^z_j \sigma^z_{j+1})\exp(-i\sum_j(h_x\sigma^x_j + h_z\sigma^z_j))$, where we take units such that $p=\hbar=1$,
depending on a triple of independent parameters $(J_z,h_x,h_z)$. KI is integrable 
for longitudinal ($h_x=0$) and transverse ($h_z=0$) fields\cite{ki}, 
and has finite parameter regions of ergodic and non-ergodic behaviors for a tilted field 
(see Fig.~1). The non-trivial integrability of a transverse kicking field, which somehow
inherits the solvable dynamics of its well-known autonomous version\cite{ti}, is quite 
remarkable since it was shown\cite{ki} that the Heisenberg dynamics can be calculated 
explicitly for observables which are bilinear in fermi operators 
$c_j = (\sigma^y_j-i\sigma^z_j)\prod_{j'}^{j'<j}\sigma^x_{j'}$ with time 
correlations decaying to the non-ergodic stationary values as 
$|C_A(t)-D_A| \sim t^{-3/2}$ \cite{ki}. 
For $D_A$ we find explicit expressions, the simplest,
\begin{equation} 
D_{\sigma^x} = \frac{{\rm max}\{|\cos(2J_z)|,|\cos(2h_x)|\}-\cos^2(2h_x)}{\sin^2(2h_x)}
\end{equation}
and $D_{M} = L D_{\sigma^x}$, for the component of spin $\sigma^x_j$, and the component of
magnetization $M=\sum_j \sigma^x_j$, respectively.

In a general situation of non-integrable KI we wish to test our theory by a numerical
experiment. We consider a line in 3d parameter space with fixed 
$J=1,h_x=1.4$ and
varying $h_z$ exhibiting all different types of dynamics: 
(a) $h_z=0$ {\em integrable},
(b) $h_z=0.4$ {\em intermediate} (non-integrable and non-ergodic), and (c) $h_z=1.4$ 
{\em ergodic} and {\em mixing}. In all cases we fix the operator $A=M$ which 
generates the perturbation of KI model
with $h_x\to h_x + (h_x^2 + h_z^2 h \cot h)\delta/h^2 + {\cal O}(\delta^2)$,
$h_z\to h_z + h_x h_z (1 - h\cot h)\delta/h^2 + {\cal O}(\delta^2)$,
where $h=\sqrt{h_x^2+h_z^2}$, and vary $L$ and $\delta$. Since 
we want the perturbation strength to be size $L$-independent we scale it 
by fixing $\delta'=\delta\sqrt{L/L_0}$ where $L_0:=24$. 
Time evolution has been computed efficiently by iterating the factored Floquet map
(in terms of 1-spin and 2-spin propagators - `quantum gates'), 
requiring $\propto L 2^L$ computer operations per iteration per initial state. 
In integrable case (a) we confirm saturation of correlations to the 
theoretical value\cite{ki} $D_M=0.485126\times L$ (Fig.~1a), as well as gaussian 
decay of fidelity (\ref{eq:fidne2}) with time-scale $\tau_{\rm ne}$ given by (\ref{eq:taune}) 
which terminates at $t\approx  t^*_{\rm ne}=\tau_{\rm ne}(\ln{\cal N})^{1/2}$ (Fig.~2a)
In non-integrable (intermediate) case (b), we find persisting non-ergodic and non-mixing 
behavior since rescaled correlation functions of typical observables $C_A(t)/\ave{A^2}$ 
relax on a short $L$-independent time scale to a non-vanishing value $D_A/\ave{A^2}$ and 
converge to TL very quickly with increasing size $L$ (Fig.~1b), but as opposed
to integrable case (a) the relaxation appears to be exponential $|C_M(t)-D_M|/L\sim 
\exp(-t/t_{\rm ave})$ with $t_{\rm ave}\approx 7.2$ (inset 1b). 
Such behavior has been observed for other two components of the magnetization $M^y,M^z$ and 
supports existence of 
intermediate dynamics observed previously in kicked t-V model \cite{ktv}. 
In Fig.~2b we confirm gaussian decay of $F(t)$ predicted (\ref{eq:taune}) 
from numerically observed value of $D_M=0.293\times L$, again up to time 
$t_{\rm ne}^*(2^L)$.
In ergodic case (c) we find fast decay of correlation functions fitting well to an 
exponential $|C_M(t)|/L \sim \exp(-t/t_{\rm mix})$, 
with $t_{\rm mix}\approx 6.0$.
Consequently we find exponential decay of $F(t)$ of eqs. 
(\ref{eq:fide},\ref{eq:taue})
using $S_M=(1/2)\sum_t C_M(t)\approx 2.54 \times L$, up to the saturation time
$t^*_{\rm e}=(1/2)\tau_{\rm e}\ln{\cal N}$ (Fig.~2c).

In conclusion, we have presented a simple theory for the stability of quantum motion with
respect to a static perturbation of the evolution operator in the limit
of Hilbert space dimension ${\cal N}\to\infty$, characterized by the fidelity measuring the
distance between time evolving states. The fidelity was expressed in terms of 
integrated time-correlation functions of the perturbing operator, 
showing that faster decay of correlations gives slower decay of fidelity, meaning
that `chaotic' dynamics is more stable in Hilbert space than `regular' one
(unless the state that one is looking at is simply related to the eigenstates of the system)!
In the two limiting cases of mixing and integrable (or more generally, non-ergodic) 
dynamics we find, respectively, exponential and gaussian decay. 
For example, our finding has strong implication for the 
stability of quantum computation with respect to static imperfections (e.g. uncontrolable
residual interaction among qubits) \cite{PZ01}. 
In other words, Eq.~(\ref{eq:taue}) is a version of the {\em fluctuation-dissipation}  
formula for the `dissipation coefficient' $1/\tau_{\rm e}$ of Eq.~(\ref{eq:fide}) which 
diverges in non-ergodic regime. 
If the system has a well defined classical limit then our formula (\ref{eq:taue})
has a clear and simple classical limit $\hbar\to 0$ too, with an integrated classical 
autocorrelation function substituting the quantum one\cite{CPW}.
We speculate that our finding is a manifestation of 
``the structural invariance''\cite{LS92} of quantum chaotic dynamics. 
Although in this paper our theory has been demonstrated in a specific 
kicked many-body problem, namely the quantum kicked Ising spin $1/2$ chain,
we should emphasize that it should be generally valid (within the time and perturbation
scales depending on the Hilbert space dimension) and thus applicable to any unitary 
evolution, in particular also to any experimentally interesting quantum dynamics.

\begin{figure}
\vspace{-8mm}
\psfig{figure=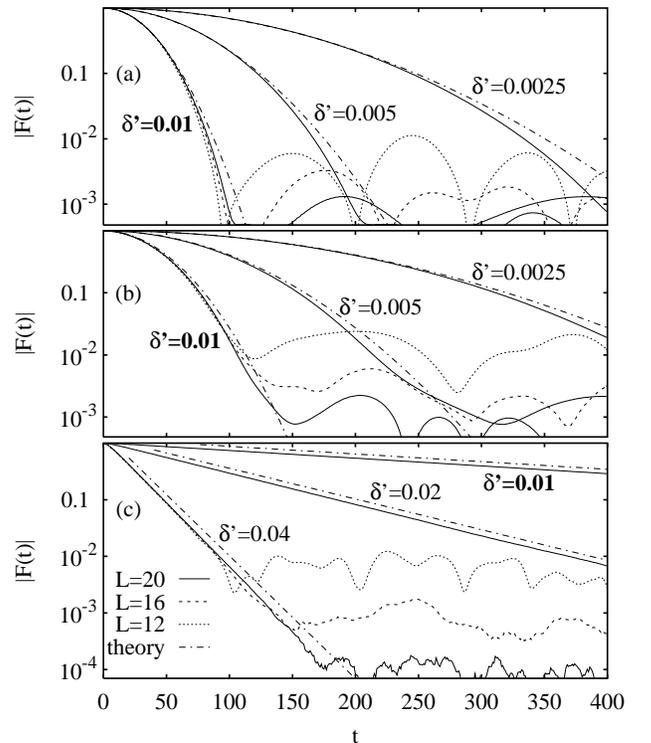,width=4.3in,angle=-90}
\caption{
Absolute fidelity $|F(t)|$ for three cases of KI: (a) integrable $h_z=0$, 
(b) intermediate $h_z=0.4$, and (c) ergodic $h_z=1.4$, for different sizes $L=20,16,12$ and
different scaled perturbations $\delta'$. Chain curves give theoretical predictions.}
\label{fig:2}
\end{figure}

The author acknowledges G.~Usaj and H.~M.~Pastawski for discussions 
in the initial stage of this
work, and T.~H.~Seligman and M.~\v Znidari\v c for very stimulating discussions and 
collaboration on related projects. The work is supported by the Ministry of 
Education, Science and Sport of Slovenia.


\begin{references}
\bibitem{foot0} We should note that working with discrete time does not
reduce generality of the results. Namely, we can simply make the limit $p\to 0$ for an
autonomous system and transform all the sums over $t$ in the expressions that follow
to integrals over time $\tau=p t$.
\bibitem{foot1} $D_A$ is a spectral weight of $C_A(t)$ at frequency $\omega=0$, whereas 
$\ave{A^2}$ is its total spectral measure.
\end{references}
\end{document}